# Model for competing pathways in protein-aggregation: role of membrane bound proteins

Youval Dar [*,+], Benjamin Bairrington [†], Daniel Cox [†,+*], Rajiv Singh [†]

[†] Department of Physics, University of California, One Shields Avenue, Davis, CA 95616
and
[+] Institute for Complex Adaptive Matter, University of California, 1 Shields Ave, Davis CA 95616

* Correspondence to: Daniel L. Cox; Department of Physics; University of California; 1 Shields Ave.; Davis, CA 95616 USA; Tel.: 415.867.4992; Fax: 530.752.4717; Email: dlcox@ucdavis.edu

Motivated by the biologically important and complex phenomena of Aβ peptide aggregation in Alzheimer's disease, we introduce a model and simulation methodology for studying protein aggregation that includes extra-cellular aggregation, aggregation on the cell-surface assisted by a membrane bound protein, and in addition, supply, clearance, production and sequestration of peptides and proteins. The model is used to produce equilibrium and kinetic-aggregation phase diagrams for aggregation onset and of reduced stable Aβ monomer concentrations due to aggregation. The methodology we implemented permits modeling of a phenomenon involving orders of magnitude differences in time scales and concentrations which can be retained in the simulation. We demonstrate how to identify ranges of parameter values that give monomer concentration depletion upon aggregation similar to that observed in Alzheimer's disease. We show how very different behavior can be obtained as reaction parameters and protein concentrations vary, and discuss the difficulty reconciling results of experiments from two vastly different concentration regimes. The latter is an important general issue in relating in-vitro and mice based experiments to humans.

## I. Introduction

Cooperative phenomena of biomolecules play many roles in living systems. One of the most important of these is protein aggregation into one-dimensional crystal-like structures called amyloids. Such a crystalline state may be the most stable phase of generic peptides, but, if unregulated, can be harmful to living organisms through unanticipated gains of toxic function. In fact, many neurodegenerative diseases, such as Alzheimer's and Mad Cow, are believed to be caused by the aggregation of certain proteins or their peptide fragments.

In the case of Alzheimer's disease (AD), one of the hallmarks of the disease is abnormal accumulation of Aβ fibrils, aggregates of small peptide fragments called Aβ mostly found in extra-cellular space. At the same time, Alzheimer's disease also gives rise to Tau-tangles, an aggregated state of the intracellular protein called Tau inside the cell. Some research suggests a connection between Aβ fibril formation and Tau-tangle formation [1,2]. In fact, there is a suggestion that the normal or cellular form of the mammalian prion protein PrP$^C$ interacts with Aβ [3-6] and may play a role in Aβ aggregate formation and its incorporation inside the cell [7-10], thus perhaps bringing it in contact with Tau proteins. Other reports, comparing mice with and without PrP$^C$, suggest that PrP$^C$ has no effect on Alzheimer's related aggregation [11,12]. There are several potential protein receptors whose internalization can bring Aβ aggregates along with them into the cell [13-15]. Several papers suggest that the Aβ might interact and alter the membrane cells secondary structure in a way that might form ion channels [16,17], we do not explore this model in this paper.

We present and simulate a protein aggregation model that incorporates a coupling of Aβ-to a generic membrane bound protein receptor we denote *PR*. Thus we present a model, where there are competing pathways for aggregation of Aβ peptides. They can aggregate by themselves in the extra-cellular region or they can aggregate on a membrane bound *PR* protein. Many membrane bound proteins are known to be internalized into the cell by a process called endocytosis, a process of transporting proteins and molecules into the cell by engulfing them in a vessel that is encapsulated into the cell. Our model includes such intake of the membrane bound proteins into the cell. This intake may also bring the Aβ aggregates inside the cell, thus providing a mechanism for interaction between Aβ and Tau. Our paper focuses on understanding the conditions under which membrane bound receptor proteins such as *PR* play a substantial role in Aβ aggregation, and when their role becomes unimportant. In particular, we show that when Aβ concentration is large, the quantitative differences between having a membrane bound receptor protein present and absent can be masked. This is important from an experimental point of view, where in mice overexpressing Aβ, the difference in aggregation between those that had *PR* and those that didn't, was found to be insignificant. The problem of mice models ability to be applicable in human is not unique to this aggregation model [18], our method shows a way to scale data from mice to humans, where the concentration of Aβ is known to be substantially smaller.

Processes like protein aggregation can be very complex and very sensitive to the conditions in the environment, with parameters such as initial concentration base production rate, and clearance rate affecting the aggregation outcome. (See Appendix for further discussion on our aggregation model parameter values.) Equations describing such reactions can be non-linear and at times difficult to solve. Assumptions on concentration and reaction parameters can be used to simplify the equations in a way that make analytical solutions or simpler simulation possible, but when, for example, a phenomenon that may appear on time scale of years involves processes on a fraction of a second scale, it is not always clear how to simplify the model while keeping is viable and representative to the phenomenon in question. In this paper we demonstrate a way to investigate a complex protein aggregation process in a way that does not require simplifying the equations prematurely and thus allowing comprehensive exploration of the process in question.

## II. Models and simulations

### A. Aggregation model

There already exist in the literature kinetic models for extra-cellular, free protein aggregation[19,20]. In several of those works, the total number of monomers or the monomer concentration is held constant. Our model expands the aggregation model to better represent in-vivo systems and more accurately, to describe possible *PR*-Aβ interaction. In addition to allowing aggregates to attach and/or nucleate on those membrane-bound *PR* proteins, we consider an open system, allowing production or supply of free monomers of Aβ from outside the cell and *PR* from within the cell. We allow extracellular clearance of Aβ monomers and aggregates, and clearance of *PR* and coupled Aβ-*PR* complexes through intracellular processes, in some aspects similar to process describe by Cisse et al [9]. We illustrate the processes in FIG. 1 and summarize the model reactions and parameter labels in Table I.

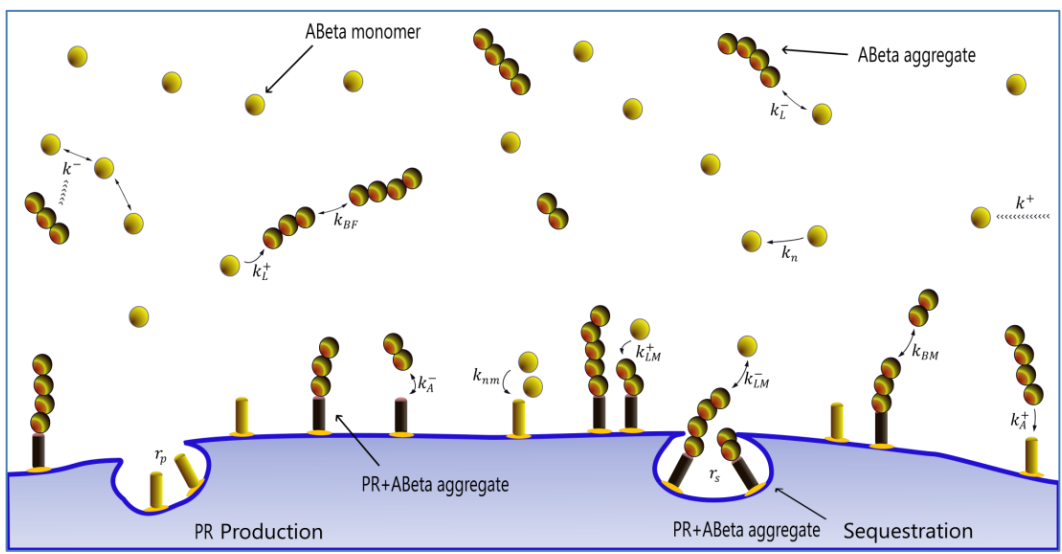

FIG. 1: Illustration of the aggregation processes in a small volume around the neuronal surface. Single sided arrows represent attachment, double sided arrows represent detachment

|     | Free aggregation | | Membrane aggregation |
| --- | --- | --- | --- |
| 1.  | $2f_1 \xrightarrow{k_n} f_2$ | 8.  | $p_0 + f_n \xrightarrow{k_A^+} p_n$ |
| 2.  | $f_1 + f_n \xrightarrow{k_L^+} f_{n+1}$ | 9.  | $p_n + f_1 \xrightarrow{k_{LM}^+} p_{n+1}$ |
| 3.  | $f_k \xrightarrow{k_{BF}} f_n + f_m$ | 10. | $p_k \xrightarrow{k_{BM}} p_n + f_m$ |
| 4.  | $f_n \xrightarrow{k_L^-} f_1 + f_{n-1}$ | 11. | $p_n \xrightarrow{k_{LM}^-} p_{n-1} + f_1$ |
| 5.  | $f_n \xrightarrow{k^-}$ | 12. | $p_n \xrightarrow{k_A^-} p_0 + f_n$ |
| 6.  | $f_1 \xrightarrow{k^-}$ | 13. | $p_0 + 2f_1 \xrightarrow{k_{nm}} p_2$ |
| 7.  | $\xrightarrow{k^+} f_1$ | 14. | $p_0 \xrightarrow{r_s}$ |
|     |    | 15. | $p_n \xrightarrow{r_s}$ |
|     |    | 16. | $\xrightarrow{r_p} p_0$ |

Table I: Reactions and reaction constants: $f_n$ is the concentration of Free Aggregates (FA) of length $n$ $(n > 1)$, $f_1$ is the free monomers concentration, $p_n$ is the concentration of Membrane-Bound Aggregates (MBA) of length $n$ and $p_0$ is the concentration of membrane proteins (*PR*). The reaction rate constants are: $k_n$ FA nucleation, $k_L^+$ monomer attachment, $k_{BF}$ FA breakage, $k_L^-$ monomer detachment, $k^-$ free monomers and aggregates clearance, $k^+$ free monomer production, $k_A^+$ single membrane protein – FA attachment, $k_{LM}^+$ monomer attachment to MBA, $k_{BM}$ MBA breakage, $k_{Lm}^-$ monomer detachment from MBA, $k_A^-$ membrane protein – free aggregates detachment, $k_{nm}$ MBA nucleation, $r_s$ membrane proteins sequestration and $r_p$ is the membrane proteins production.

Our aggregation model is described by the set of reactions presented in Table I. The model includes nucleation of free aggregates from two free Aβ monomers (1). Once nucleated, aggregates can grow longer or get shorter by attachment or detachment of one Aβ monomer (2, 4), attachment from one side, detachment from both. When long enough, aggregates can break (3). We include Aβ production (6,7) and clearance of Aβ and free aggregates (5). Similar interactions are introduced for the membrane bound proteins (*PR*). Free aggregates can attach to a membrane protein (*PR*) to create a membrane bound aggregate (8). A single Aβ monomer can attach or detach to a membrane aggregate (9, 11). Membrane aggregates can break in two, a free aggregate and a membrane-bound one (10). Membrane bound aggregates can break free from the membrane; break from the *PR* (12). The *PR* can also induce aggregate nucleation, a process where two Aβ monomers join to a membrane *PR* to form a membrane bound aggregate (13). All reactions are assumed to be taking place in a reaction-volume outside the cell. The model includes production and clearance of *PR* (14,15, 16) into the reaction-volume, which we call production and sequestration. When *PR* is being sequestered back into the cell while attached to an aggregate, the aggregate is also pulled into the cell and removed from the reaction-volume. We assume that the net flux of monomers in a given reaction volume is zero. We also assume that production and clearance of monomers is balanced when no aggregation occurs, so $k^+$ and $k^-$ are dependent as well as $r_p$ and $r_s$. The motivation for the model parameters values, used in our simulation are described in the appendix.

### B.   Phase Diagram

Our aggregation model has dynamical phases that can be characterized by different physical properties, such as aggregate concentration and sizes, the concentration of monomers and by the dynamical time dependent patterns of monomers and aggregates. For example, there are phases: (i) *Subcritical (S)* with essentially no transient aggregation and unchanged monomer concentration changes. (ii) *Aggregated (A)* steady-state aggregation with monomer concentration reduced by a definite fraction. (iii) *Transient Subcritical (TS)* with observable levels of transient aggregates.

Our goal is to explore the range of dynamical phases in the phase space determined by species concentrations and reaction parameters.  We have developed a computational algorithm to efficiently determine these phase diagrams. The details of the algorithm are provided in the appendix. In particular, we create phase diagrams for state *A* state, which represents a state with Alzheimer's Disease (AD).

### C.   Hybrid stochastic-deterministic study

Due to the time scale differences between monomer production and the onset of AD we use a combination of stochastic and deterministic methods to computationally simulate the process and obtain the phase diagram for the system. We have used several sources [12,19] to guide us in choosing reaction parameter values, *PR* concentrations, and Aβ concentration but we modified those values according to the physical state we were exploring.

The reactions in Table I give rise to a large set of nonlinear-coupled ordinary differential equations. In the model, the Aβ and *PR* concentrations can be very small and vary widely. Moreover, there can be many orders of magnitude difference in the values of reaction constants. An analytic solution that requires linearization and simplification of the equations is highly problematic if one do not want to risk losing important physical characteristics of the model. We will, instead, solve these equations by a predominantly stochastic numerical method, appropriate since some species have relatively small number of particles.

The stochastic simulation of a chemical reactions set is based upon the probabilistic propensity of the reactions to occur and uses number of particles in a reaction volume, not by concentrations. In our model, the monomer production and clearance reactions are far more likely to occur than any other reaction. Thus, a purely stochastic simulation will spend most of its time doing the trivial task of bringing monomers in and out of the system. Since the aggregation time can be of the order of years at low concentrations, including monomer supply and clearance as stochastic variables will make exploration of this model impractical. We therefore use a hybrid approach with aggregation and aggregate clearance reactions treated stochastically via Gillespie's algorithm[21], and monomer creation and clearance treated deterministically. The deterministic portion is designed in such a way that in the absence of aggregation one gets homeostasis of the monomers at values compatible with experimentally estimated concentrations.

In the Gillespie method we use the probability distribution or propensity of each reaction to occur and stochastically choose the time till next reaction $dt_{Gillespie}$ and reaction type $(\mu)$. The Aβ production parameter $k^+$ and PR production parameter $r_p$, were set according to $k^-$ and $r_s$, the clearance and sequestration parameters, in such way that a lack of aggregation will produce the desired stable concentration, $k^+ = [A\beta](t=0)k^-$ and $r_p = [PR](t=0)r_s$. We choose initial aggregate nucleation rate on the membrane to be comparable to free nucleation rate and use $n_c = 2$ (dimers) as the critical aggregate nucleus size. All concentrations were converted to number of particles. All reaction parameters were converted to propensities according to the Gillespie method[21]. A concentration of $x_i$ was converted to number of particles using $X_i = vN_A x_i$, where $v$ is the reaction volume in liters and $N_A$ Avogadro's number. The Gillespie reaction parameters conversion method for the reaction in Table I is described in equations (1) to (3) using $V \equiv vN_A$. Quantities $x$ and $y$ in equations (1) to (3) are generic symbols for concentration of different species.

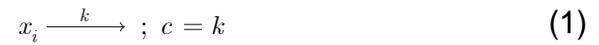

$$x_i \xrightarrow{k} \quad ; \quad c = k \tag{1}$$

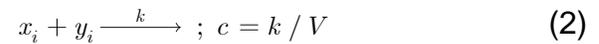

$$x_i + y_i \xrightarrow{k} \quad ; \quad c = k/V \tag{2}$$

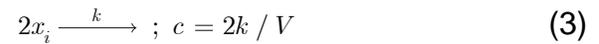

$$2x_i \xrightarrow{k} \quad ; \quad c = 2k/V \tag{3}$$

The propensity for each reaction type to happen is $c * h$ where $c$ is the converted reaction constant, and $h$ (Table II) is the number of ways a particular reaction can happen. In Table II, we denote the number of PR that are not attached to aggregates as $x_0(t)$, the number of PR that are attached to an aggregates of length $n$ as $x_n(t)$, the number of free Aβ monomers as $y_1(t)$ and the number of free aggregates of length $n$ as $y_n(t)$.

| | |
|---|---|
| $h_1 = y_1(y_1 - 1)/2$ | $h_9 = x_n y_1$ |
| $h_2 = y_1 y_n$ | $h_{10} = x_k(k-2)\ [n \geq 1, m > 1]$ |
| $h_3 = y_k(k-3)\ [n, m > 1]$ | $h_{11} = x_n$ |
| $h_4 = 2y_n\ [n > 2]$ | $h_{12} = x_n$ |
| $h_8 = x_0 y_n$ | $h_{13} = x_0 y_1(y_1 - 1)/2$ |

Table II: Summary of $h_i$, the number of ways reaction $i$ can from Table I can occur, according to the number of particles involved in the reaction. $y_n$ is the number of free aggregates of length $n$. $x_k$ is the number of membrane bound aggregates of length $k$. $x_0$ is the number of membrane bound proteins without aggregates.

Noting the propensity for reaction $\mu$ as $a(\mu) = h_\mu c_\mu$ with $\mu \in (1, M)$, and setting $a_0 = \sum_{\nu=1}^{M} h_\nu c_\nu$, using a random variable $r_1$, the time until the next reaction is

$$dt_{Gillespie} = \frac{1}{a_0} \ln(\frac{1}{r_1}) \quad (4)$$

The type of reaction is chosen based on the aggregates population, using the second random variable $r_2$ and the equation

$$\sum_{\nu=1}^{\mu-1} \frac{a_\nu}{a_0} < r_2 < \sum_{\nu=1}^{\mu} \frac{a_\nu}{a_0} \quad (5)$$

During the waiting time $dt_{Gillespie}$ between the reactions, the monomers and only the monomers concentration, $y_1$ and $x_0$ continues to evolve deterministically through their production and clearance. This is governed by the equation

$$\frac{dx_0(t)}{dt} = r_p - x_0(t)r_s, \quad (6)$$

With a solution

$$x_0(t) = \frac{r_p}{r_s}(1 - e^{-r_s t}) + x_0(0) e^{-r_s t}. \quad (7)$$

Over the time $dt_{Gillespie}$ it leads to the change

$$dx_0 = \left(\frac{r_p}{r_s} - x_0(t)\right)(1 - e^{-r_s dt_{Gillespie}}). \quad (8)$$

The integer part of this change is added to $x_0$. The non-integer fraction is kept in a buffer and used in the next time-step. Equations (6,7,8) also apply to $y_1$ with its corresponding production and clearance constants.

In the Gillespie method, after each reaction, the number of aggregates that were affected by the reaction is changed.

All relevant propensities are updated using those new numbers then the next time step and reaction are chosen using new random numbers.

The aggregation onset time was evaluated in two different ways, the time till we have sustainable aggregates levels, aggregation and the time till the rate of the aggregation changes by a large factor over a set time. Aggregation transition time is the time between aggregation onset and the end stable state, when one exists.

We tested our program successfully by comparing our aggregation results for no surface receptors and fixed monomer concentration to those presented in papers by Knowles et al[19] and Kunes et al[20].

### III. Results

We start by exploring states where the aggregation process can be sensitive to *PR* and other states where the *PR* impact can be missed. We explore how the kinetic diagram of the aggregation process may look like and how changing reaction parameter and nucleation characteristics might affect the aggregation kinetics. We then created phase diagrams for several concentration and reaction parameters. Table IV contains the reaction parameters values for all figures.

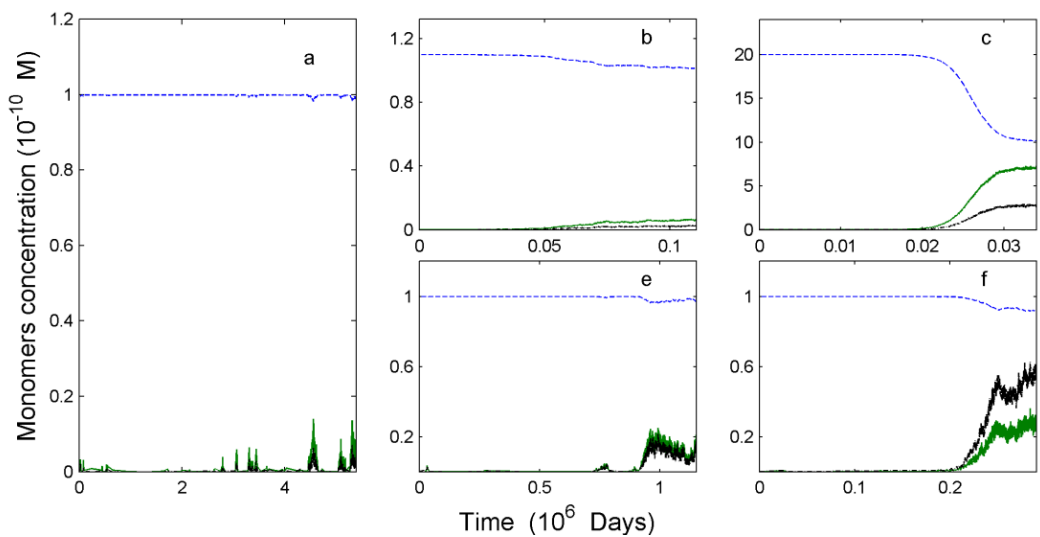

FIG. 2: Critical aggregation phase transition due to changes in concentration. (a) no sustainable aggregation. (b,c) Critical behavior due to $[A\beta]$. (e,f) Critical behavior due to $[PR]$. Blue (dash) line – Aβ monomers. Black (dash-dot) line – *PR*/membrane bound aggregates. Green (solid) line – free aggregates. Reaction parameters are in Table IV. (a-c) initial $[A\beta]$ are 0.1nM, 0.11nM and 0.2nM respectively and $[PR]$ are 2nM, 4nM and 10nM. In Plots (a), (e) and (f) The Aggregates concentration levels (black and green lines) are multiplied by 10.

FIG. 2 shows how sensitive the critical aggregation state can be. FIG. 2 (a) shows a state in the *Subcritical* phase; random aggregate concentration blips arise but fall back down to zero promptly. In FIG. 2 (b) a change of 0.01nM in the Aβ concentration initiates aggregation *(A phase)*. In FIG. 2 (c) we further increase the Aβ concentration. As expected, increase in the Aβ concentration shortens the aggregation onset time as well as increases the resultant aggregate concentration. Though quantifying onset time, aggregation levels and whether the aggregation levels are stable or not is not trivial and it depends on other model parameters

Decreasing the Aβ concentration also increases the variation in the onset time. FIG. 2 (e) shows how changing the *PR* concentration can cause aggregation and in FIG. 2 (f) we see how higher concentration shortens the onset time and increases the aggregate levels.

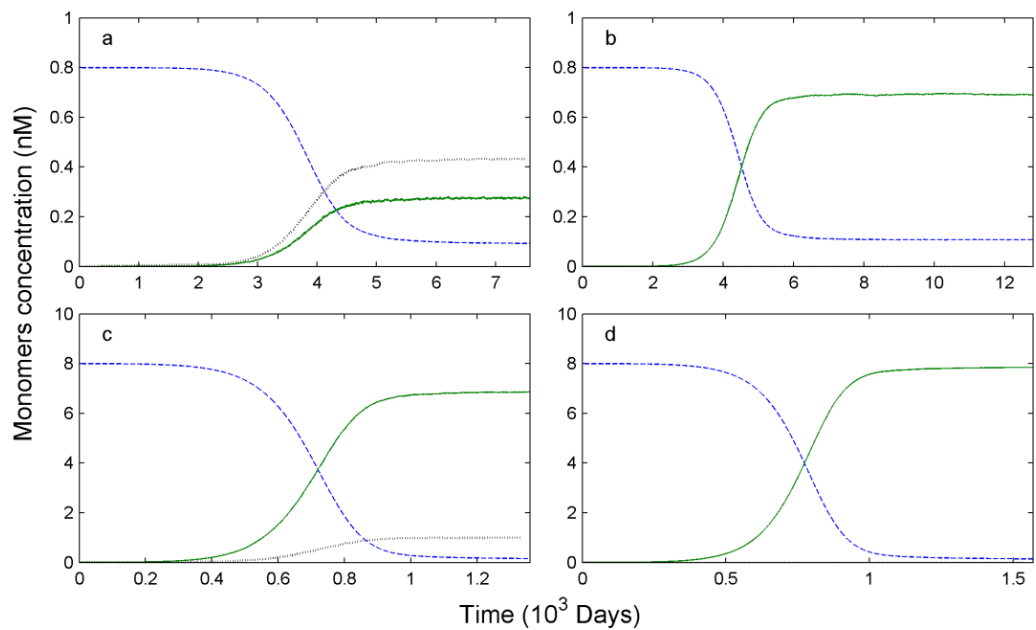

FIG. 3: *PR* effect on the aggregation process. Blue (dash) $[A\beta]$, Green (solid) [Free Agg.], Black (dot) [Membrane Agg.]. (a): Stable aggregates and $[A\beta]$ levels, $[PR]$ is 8nM, (b): Without *PR*, (c): High Aβ concentration, $[PR]$ is 0.8nM and (d): Without *PR*.

In FIG. 3 we explore the effect of eliminating *PR* from the reaction. FIG. 3 (a) shows a situation where we have both free aggregation and membrane bound aggregates; with time we see a reduction in Aβ monomer concentration level. FIG. 3 (b) shows how eliminating *PR* from the system hardly changes the total aggregate concentration, and only slightly affects the aggregation onset time. FIG. 3 (c) and FIG. 3 (d) show a situation where the *PR* concentration is small compared to Aβ concentration. The system has many more free aggregates and thus elimination of *PR* and membrane aggregation has a small effect on the total aggregate concentration or the aggregation onset time. Both (c) and (d) show that eliminating the membrane aggregates increases the free aggregates. In (c) and (d) the Aβ monomers are nearly depleted after aggregation is stabilized. The Aβ level that is left at the end of the process is sensitive to the monomer attachment and detachment rates. High attachment and detachment rates can cause near depletion of Aβ monomers from the system.

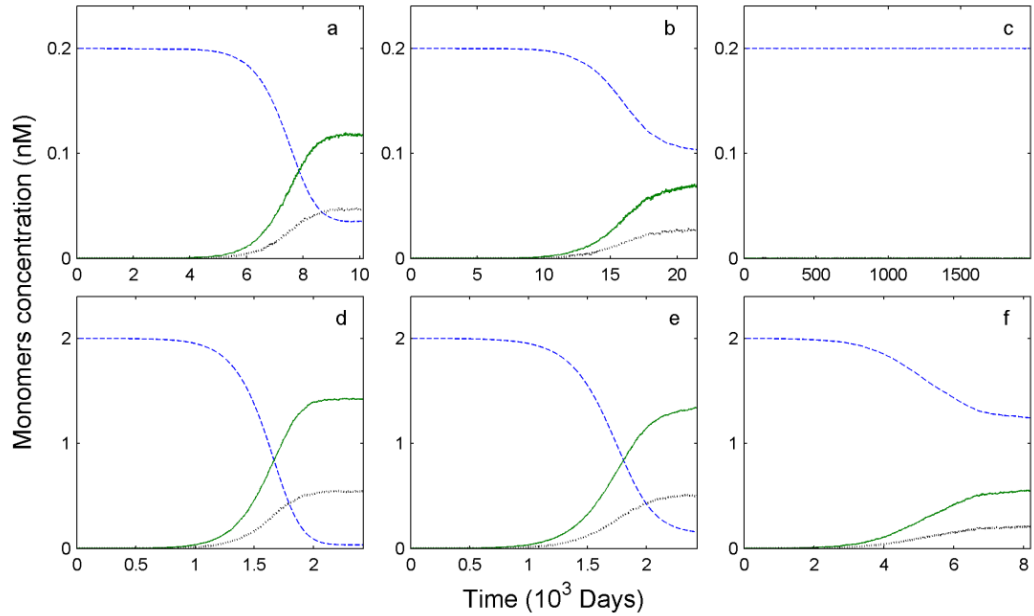

FIG. 4: The effect of clearance rate on the aggregation kinetics. In all plots $k^- = r_s$ (clearance = sequestration). (a,d): $k^- = 0$. (b,e): $k^- = 1 \times 10^{-8} s^{-1}$, (c,f): $k^- = 5 \times 10^{-8} s^{-1}$. In plots (a-c) $[A\beta] = 0.2nM$, $[PR] = 2nM$. In plots (d-f) $[A\beta] = [PR] = 2nM$. Blue (dash) $[A\beta]$, Green (solid) [Free Agg.], Black (dot) [Membrane Agg.]

FIG. 4 demonstrates time evolution of aggregate accumulation and how this evolution can be affected by varying the reaction parameters $k^-$ and $r_s$. We demonstrate how the increase of the clearance rate changes the characteristics of the process. The concentrations in (a-c) are representative of those in humans, with higher *PR* and lower Aβ levels. In (d-f) the Aβ concentration is elevated, to represent experimental conditions in mice. The aggregate levels change as the production and clearance rates change. In (c), the increased clearance rates prevent aggregation formation, in sharp contrast to the same rates at a higher Aβ concentration levels. The clearance rates had some influence on the onset time, though some variation is due to the stochastic nature of the simulation.

When changing the relative clearance rates of aggregates and monomers, a slower clearance of aggregates and faster production of Aβ monomer causes accumulation of aggregates in the system. This effect is more pronounced in the situation with elevated Aβ monomer concentration.

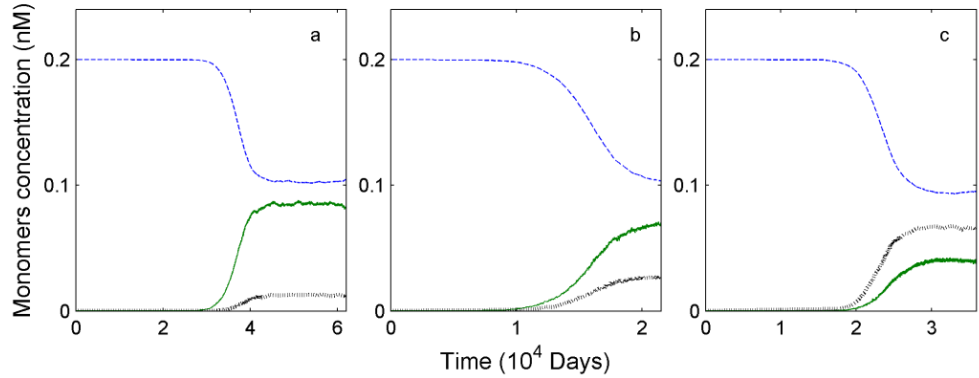

FIG. 5: Effect on aggregation kinetics of aggregate – membrane interaction propensity that is aggregate length dependent. In all plots $k^- = r_s = 10^{-8} s^{-1}, [A\beta] = 0.2nM$ and $[PR] = 2nM$. (a): $k_A^+(n) = k_A^+ / n$ for aggregates of length larger than 10, (b): length independent $k_A^+$ and (c): $k_A^+(n) = nk_A^+$ for aggregates of length larger than 10. Blue (dash) [Aβ], Green (solid) [Free Agg.], Black (dot) [Membrane Agg.]

This method allows exploration of aggregation model nuances; in FIG. 5 we explore how a membrane interaction propensity that depends upon oligomer size might affect the aggregation kinetics. To simulate low propensity of large aggregates for membrane interaction, we change the propensity to attach to the membrane protein, so that for aggregate length n>10 we reduce the attachment rate parameter to $k_A^+ \rightarrow k_A^+ / n$ FIG. 5 (a). One might envision a situation where longer aggregates are slower to clear the system and more likely to get stuck on membrane proteins in a way that allows more time for those long aggregates to form bonds. To explore this scenario, we change the propensity of aggregates to attach to the membrane so that for aggregate length n>10 we have an attachment rate parameter of $k_A^+ \rightarrow nk_A^+$ FIG. 5(c). The aggregation onset time appears shorter when there is no attachment length dependence FIG. 5 (b), slightly longer when we favor longer aggregates attachment and slightly longer when we penalize longer aggregates attachment. The difference in the onset time is not large and might be a result of the stochasticity of the process. The final concentration of Aβ monomers also appears to be similar in all plots. However, the amount of free vs. membrane aggregates is very different. When long aggregate attachment is favored, most aggregates end up attached to the membrane, while most aggregates are free of the membrane when longer aggregate attachment to the *PR* proteins is penalized.

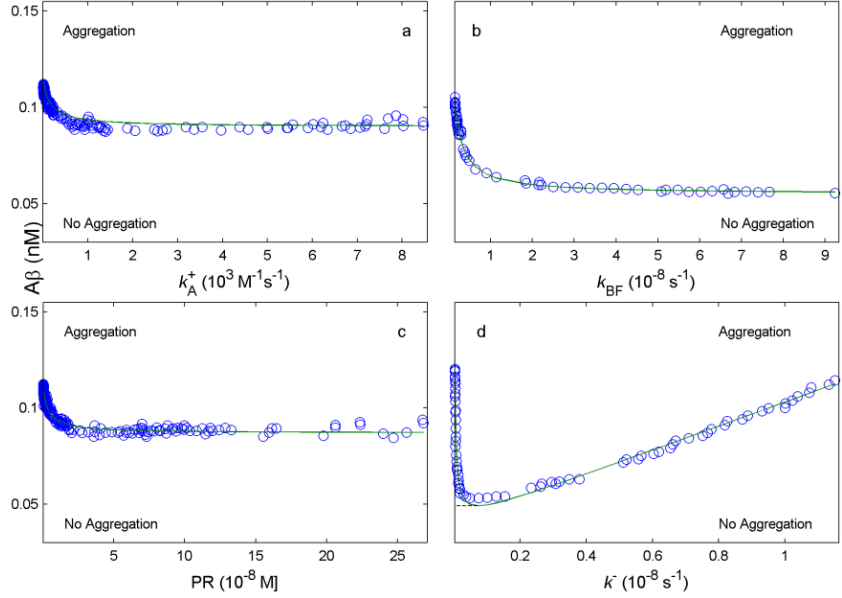

FIG. 6: Critical aggregation transition phase line. (a): $[A\beta]$ vs. $[k_A^+]$, (b): $[A\beta]$ vs. $[k_{BF}]$, (c): $[A\beta]$ vs. $[PR]$, (d): $[A\beta]$ vs. $[k^-]$. Above the phase line aggregates are formed, under the line there is no sustainable aggregation. Reaction parameters in Table IV

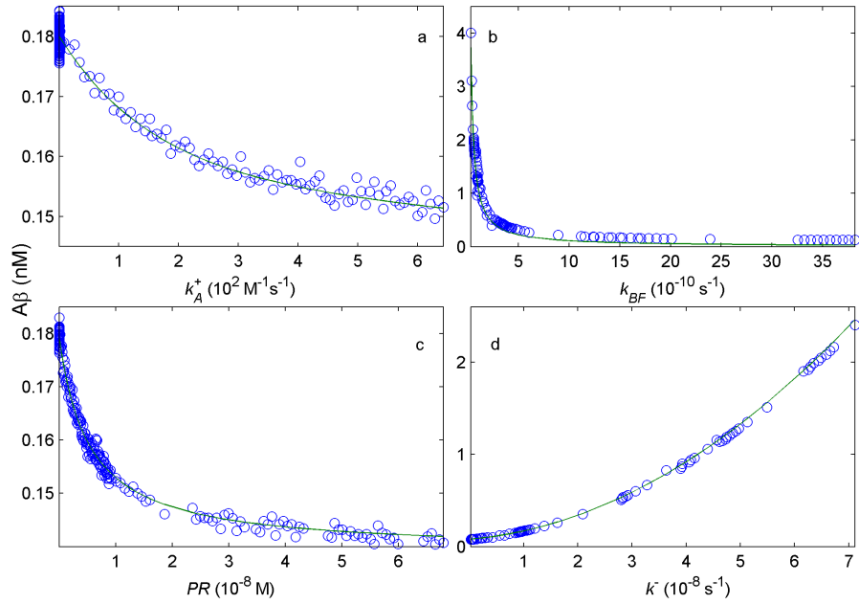

FIG. 7: 40% $[A\beta]$ stable level reduction phase line. (a): $[A\beta]$ vs. $[k_A^+]$, (b): $[A\beta]$ vs. $[k_{BF}]$, (c): $[A\beta]$ vs. $[PR]$, (d): $[A\beta]$ vs. $[k^-]$. Above the phase line $[A\beta]$ higher than 40%, under the phase line it is lower. Reaction parameters in Table IV

In FIG. 6 and FIG. 7 demonstrate investigation of two phenomenon's, critical aggregation and a process of aggregation that ends at reduced stable Aβ levels, we set initial [Aβ] and run the model, the [Aβ] is not held at the initial level, but is allowed to fluctuate according to the model. Those figures demonstrate phase diagrams for $[A\beta]$ vs. $[k_A^+]$ (a), $[A\beta]$ vs. $[k_{BF}]$ (b), $[A\beta]$ vs. $[PR]$ (c) and $[A\beta]$ vs. $[k^-]$ (d) for the two example process, 40% reduction and stable level of [Aβ] state and the critical aggregation state. In addition to showing the relation between the reaction parameters, the phase diagrams reveals non-trivial relations of the phase boundary, in that aggregation no-aggregation boundary can be at different monomers concentration when following different reactions parameters, or that a particular phenomenon might not realistically exist for some values of reaction parameters. The phase diagrams also show that there is some initial Aβ concentration above which aggregation always happens and above which the Aβ concentration ratio observed between healthy human and AD patients might not be attainable. Simulations with different reaction parameters produced different phase curves. We fitted different curves types to the phase diagrams. We denote rational fit curves with *RAT*, power fit with *POWER* and exponential fit with *EXP*. RAT, POWER and EXP curves are defined in equations (9),(10), (11) and (12). The parameters for the different figures fitted curves are in Table III.

$$y_{RAT} = \frac{p_1 x^2 + p_2 x + p_3}{q_0 x^2 + q_1 x + q_2} \qquad (9)$$

$$y_{POWER} = p_1 x^{q_1} + p_2 \qquad (10)$$

$$y_{POLY} = p_1 x^2 + p_2 x + p_3 \qquad (11)$$

$$y_{EXP} = p_1 \exp(p2 \cdot x) + q_1 \exp(q_2 \cdot x) \qquad (12)$$

|  | Fit function | $p_1$ | $p_2$ | $p_3$ | $q_0$ | $q_1$ | $q_2$ |
|---|---|---|---|---|---|---|---|
| FIG. 6(a) | RAT | 0 | 0.09 | 0.19 | 0 | 1 | 1.66 |
| FIG. 6(b) | RAT | 0 | 0.055 | 0.134 | 0 | 1 | 0.632 |
| FIG. 6(c) | RAT | 0 | 0.087 | 0.437 | 0 | 1 | 3.962 |
| FIG. 6(d) | RAT | 2488 | 1580 | 17 | 1 | 39540 | 78 |
| FIG. 7(a) | EXP | 0.23 | -0.65 |  |  | 1.57 | -0.006 |
| FIG. 7(b) | POWER | 11 | 0.0 |  |  | -1 |  |
| FIG. 7(c) | RAT | 0 | 1.39 | 0.95 | 0 | 1 | 0.53 |
| FIG. 7(d) | POLY | 0.04 | 005 | 0.08 |  |  |  |

Table III: Fitting parameters for equations (9,10,11), FIG. 6 and FIG. 7

In FIG. 6 we evaluate the critical aggregation by looking at the average amount of aggregate over time and by the aggregation sustainability. Two interesting phenomenon can be observed, the first, since our model does not enforce total number of monomer in the reaction volume. Since the clearance and creation of monomers are related to each other and to the initial Aβ levels, and the clearance rate of Aβ monomers is the same as clearance rate of free aggregates, at a slow monomers creation and clearance rate the clearance of aggregates can reduce significantly the number of Aβ in the system and may create non-monotonic behavior as observed in FIG. 6.(d) .Another observation is that when the Aβ level is low and the *PR* is high, the stochastic variability is very large, aggregation become *Subcritical,* where no aggregation occurs, *Transient Subcritical,* where aggregation forms but does not last in some osculating manor, and the critical aggregation line becomes blurred and hard to determine. A particular range of Aβ concentration, production rate and corresponding sequestration rate is needed to make for a clear distinct appearance of the critical aggregation line. This phenomenon is very visible in FIG. 6.(d) but when producing FIG. 6 (a) and (b) that tendency was visible and was more pronounced at higher values of $k^-$ and $k_{BF}$.

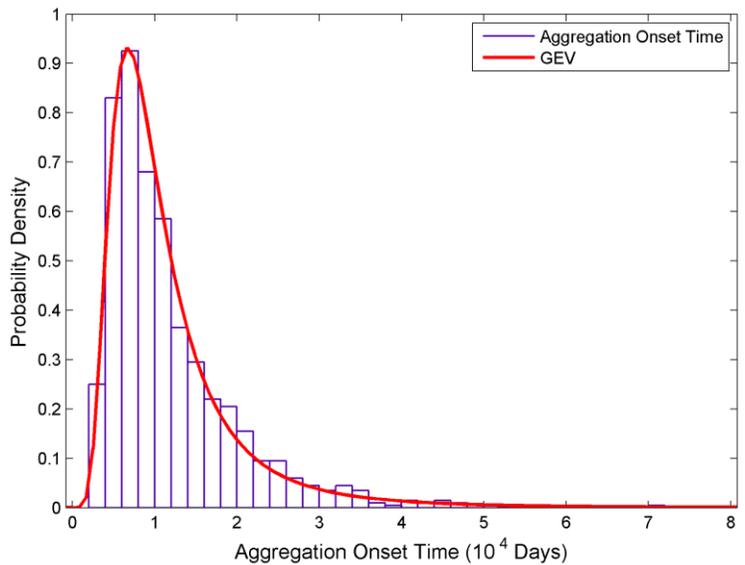

FIG. 8: Onset time distribution histogram with a generalized extreme value (GEV) distribution fit. GEV probability function parameters: k (Shape) = 0.3165, σ (Scale) = 0.4129 and μ (Location) = 0.7790, at 95% confidence level. GEV typically used for assessing various financial risks and phenomenon with extreme deviation from the median of event probability distribution.

The stochastic variability of the aggregation onset time increased as the number of particles in the simulation volume decreased. FIG. 8 shows the onset time distribution for the simulation with parameters adapted for humans, using the parameters in the last column in Table IV and with 1,000 iterations.

| Reaction parameters | Critical aggregation FIG. 2 | | PR indifference FIG. 3 | | Production rate and aggregation kinetic FIG. 4 | | Critical aggregation FIG. 6 | %40 reduction in Aβ Levels FIG. 7, 8 |
|---|---|---|---|---|---|---|---|---|
| | a-c | e,f | a,b | c,d | a-c | d-f | State point | State point |
| $k_L^+$ $(M^{-1}s^{-1})$ | 1500 | 1500 | 1500 | 1500 | 1500 | 1500 | 1500 | 1500 |
| $k_L^-$ $(10^{-8}\ s^{-1})$ | 3 | 3 | 3 | 3 | 3 | 3 | 3 | 3 |
| $k_{LM}^+$ $(M^{-1}s^{-1})$ | 1500 | 1500 | 1500 | 1500 | 1500 | 1500 | 1500 | 1500 |
| $k_{LM}^-$ $(10^{-8}\ s^{-1})$ | 3 | 3 | 3 | 3 | 3 | 3 | 3 | 3 |
| $k_A^+$ $(M^{-1}s^{-1})$ | 100 | 100 | 100 | 100 | 100 | 100 | 100 | 100 |
| $k_A^+$ $(10^{-8}\ s^{-1})$ | 50 | 50 | 50 | 50 | 50 | 50 | 50 | 50 |
| $k_n$ $(10^{-5}\ M^{-1}s^{-1})$ | 9.5 | 9.5 | 9.5 | 9.5 | 9.5 | 9.5 | 9.5 | 9.5 |
| $k_{BF}$ $(10^{-8}\ s^{-1})$ | 0.15 | 0.15 | 0.15 | 0.15 | 0.15 | 0.15 | 0.15 | 0.15 |
| $k_{BM}$ $(10^{-8}\ s^{-1})$ | 0.15 | 0.15 | 0.15 | 0.15 | 0.15 | 0.15 | 0.15 | 0.15 |
| $A\beta\ (nM)$ | Vary | 0.1 | 0.8 | 0.8 | 0.2 | 2 | 0.1 | 0.17 |
| $PR\ (nM)$ | 2 | Vary | Vary | Vary | 2 | 2 | 2 | 2 |
| $r_s$ $(10^{-8}\ s^{-1})$ | 1 | 1 | 1 | 1 | Vary | Vary | 1 | 1 |
| $k^-$ $(10^{-8}\ s^{-1})$ | 1 | 1 | 1 | 1 | Vary | Vary | 1 | 1 |

Table IV: Reaction parameter values for the simulations presented in this paper. See Appendix for experimental values.

## IV. Discussion

In this work, we have developed a model for aggregation that includes extra-cellular and membrane-assisted aggregation as well as mechanisms for supply and clearance of monomers and aggregates. Using a hybrid stochastic-deterministic method we were able simulated complex aggregation process without the need to simplify it, keeping non-linear relations and parameters with orders on magnitude values differences. The Gillespie method was used for the aggregation reactions and a deterministic method for the monomers supply and clearance. Examining the aggregation kinetics and the effects of different reaction parameters on it, we identified critical conditions for onset of aggregation, conditions that produced stable levels of aggregates and Aβ; and conditions where once aggregation started the amount of aggregates increased with time. For such states we obtained appropriate phase diagrams.

In the context of phenomenon that might not be observed in mice but might still be important in human, we observed that several of the reaction parameters and concentration can have a significant effect on the aggregation onset time, aggregate levels, and on whether aggregation will occur or not. In some of the states that we explored, we observed that the initial Aβ and *PR* concentration levels and the corresponding monomer production rates can have significant effect on the aggregation process, while at high levels of Aβ we observed that the *PR* contribution to the total amount of aggregates can be small. The critical line between states with no time averaged aggregates and states with steady state aggregates, as well as some states of stable aggregation were inaccessible under those high [Aβ] conditions. At lower levels of Aβ and *PR* we observed high sensitivity of the aggregation process to concentration changes.

We observed that clearance and production rates can dramatically change the characteristic of the aggregation kinetics (FIG. 4). Clearance and production rate can affect the levels of monomers and aggregates, whether the level of aggregates is stable or continuously rising, the aggregation onset time, and can even determine if aggregation will happen at all. Attachment and detachment rates of monomers to aggregates and the ratio between them also affected the end level of Aβ monomers and the aggregation transition time. A high attachment-to-detachment rate ratio meant that almost all monomers would attach to some aggregate as soon as they enter the system. High values of those rates meant that the aggregation transition time was shorter. The ratio between the free nucleation and the membrane assisted nucleation was also important to the role of the *PR*. In this paper we only show results where the free and membrane assisted nucleation were comparable. We also observed that the smaller the number of particles in our system, the greater the fluctuation in aggregation onset time and the smaller the maximum size of aggregate.

The number of particles in the system is a consequence of the concentration and the reaction volume. The reaction volume should represent the physical dimension of the brain and the relevant neuronal environment where the aggregation takes place. In our simulations, the number of particles in the system was in the range of hundred thousand to millions, which we believe are reasonable numbers for in-vivo simulation. With those particle numbers and the reaction parameters we used, the maximum aggregates length we needed to use was 500–800.

| Reported Aβ levels | |
|---|---|
| Heathy human CSF Aβ$_{42}$ | 700(250) pg/mL |
| Human with AD CSF Aβ$_{42}$ | 451(178) pg/mL |
| Human with AD Soluble Aβ | 8.6±2.1 pmol/gr |
| Mice Soluble Aβ | 1629 ± 380 pmol/gr |
| PrP$^C$ cell-surface expression | |
| Human 6H4/Blood PLT Cell | 619 ± 167 |
| Mouse 6H4 /Blood PLT Cell | 5 ± 3 |

Table V: mean (standard deviation) of CSF Aβ$_{42}$ in healthy and human with AD [22,23]. Soluble Aβ$_{40}$ and Aβ$_{42}$ levels as reported for a APP/PS1 mouse model of AD and Humans [24]. Expression of cell-surface PrP$^C$, measured as the number of anti-prion mAb 6H4 molecules bound per blood cell in human and mouse platelets (PLT) [25,26]

In human AD it's reported that Aβ levels in healthy people are about 40% higher than that of AD patients[22,23,27] and that the levels of aggregates in AD patients does not increase with time[28-30]. We examined a phase state that has those features and found that this state is not far above the critical aggregation state in terms of Aβ levels (In Table III the parameters for FIG. 6 and FIG. 7 show 0.07nM difference between those states). This means that our model supports a scenario in which increased level of Aβ can trigger an aggregation process that ends in the experimentally observed Aβ and aggregate levels state.

There are conditions in mice experiments that are different than those of human and that can be accounted for in our model. The Aβ concentration can be many-fold larger in mice[24] (Table V), and clearance and production rates can be very different than those in humans. Indeed it appears that aggregate levels increase with time[31] ,and possibly the secondary structure of the aggregates is suspected to be different[24]. We showed how the formation of Aβ aggregate can be, both sensitive and indifferent to *PR* presence depending on parameters. FIG. 2 demonstrated how small increases in *PR* levels can initiate aggregation. On the other hand, FIG. 3 shows how the effect of the *PR* can be missed, either because the total amount of aggregates in the system remain the same, or due to much higher Aβ levels that result in free aggregation dominating the process. FIG. 6 reveals that above some Aβ concentration level aggregation will always happen, and that we might lose sensitivity to other reaction parameters effects on aggregation.

The kinetic diagram (FIG. 4) can be an important link between experiments and our model, since the kinetic diagram can be constructed experimentally by measuring monomer and aggregate levels. Additional links between experiments and our model can be the length of aggregates found in mice or humans, the fluctuation in aggregation onset time, the aggregation transition time and the relevant reaction volume in the brain where the aggregation takes place. The kinetic diagram can also be used to explore nuances in the aggregation model as demonstrated in FIG. 5, where we looked at the dependence of membrane attachment propensity on oligomer size.

Our model includes sequestration of *PR*-Aβ aggregates; it can give estimated amounts of Aβ aggregates in the cell as a function of time. This information can be very useful when exploring the relations between the Alzheimer's Aβ fibrils and Tau tangles and the mechanisms by which the Aβ fibrils could promote Tau tangle formation.

The phase diagrams are new useful tools produced by our model and simulation. A tool to compare and interpret phenomenon that happens in different environments. Once the phase diagrams for a particular phase state in humans are known, they can give insight into the concentration levels and reaction parameter values that should produce a similar state in mice, and help design and interpret experiments. The phase diagrams can also provide a link between different mice experiments and may help gain knowledge on the values of reaction parameters, including breakage rates, attachment rates, nucleation rates and so on, that are not known.

The median of the aggregation onset time, the time in which the aggregation rate increases significantly and after which the system contains sustainable elevated amount of aggregates, as describe in right skewed FIG. 8, and FIG. 5 (b) is about 26 years. The aggregation transition time, FIG. 5, have about the same scale, which make the time scale of 54 years FIG. 5 (b) or 69 years FIG. 5 (c), years to get to a stable AD like aggregation levels

The hybrid stochastic-deterministic method, provides a powerful ability to simulate very different reaction parameters scales. Perhaps an improvement to this method can made by switching to a pure stochastic model when the production and clearance time scales becomes similar to the Gillespie time steps. In particular in phase exploration such as in Fig 6 and 7, some behavior might have been affected hybrid stochastic-deterministic model.

### V. Acknowledgement:

We acknowledge useful conversations with Gil Rabinovici. This work was supported in part by the International Institute for Complex Adaptive Matter, US National Science Foundation Grant (YD) DMR-0844115, and by US National Science Foundation Grant DMR-1207624 (DLC and RRPS).

### VI. Appendix

#### A. Reaction and simulation parameters

The reaction parameters in our model are not well known, but some of the aggregation physical traits are known to some degree. The concentrations of Aβ and *PR*, though not uniform in the brain, are known at a coarse grained level. We know the AD aggregation time characteristics, decades for aggregation onset time and several years for aggregation transition. We also wanted to reproduce the observed reduction in Aβ monomers levels and the stable or slowly growing amount of aggregates in AD patients. We considered equal, free and membrane, initial nucleation rates and assumed that the concentration of Aβ is lower than that of *PR*. The volume we used in our simulation was $10^{15}nm^3$, the volume unit is such that concentration of 2nM gives about $10^6$ particles. This volume represents the region between neurons, where the aggregation takes place. A human brain volume is about 1.2 liter and contains about 0.14 liter of cerebral fluid. There are about 85 billion neurons in a human brain. This implies that there are about 600 neurons in and around our reaction volume.

The larger the number of particles in a system, the larger the maximum length of aggregate we need in the simulation. The reaction parameters and clearance rate also affect the aggregate size distribution and maximum aggregate size. We set this at 500 and monitored the simulation, to ensure that there were no aggregates exceeding this length.

Aβ production is such that about 7% of total number of particles are being produced every hour[32,33]. In the absence of aggregation, the production and clearance are assumed to be at homeostasis. This gives us $k^+ \approx 2[s^{-1}]$ and $k^- = 2 \times 10^{-5}[s^{-1}]$ which implies a residence time of the order of several hours $1/k^-$. We took the residence time of a *PR* on the cell membrane to be about 11 hours (40000s). This gives $r_s = 40000^{-1}[s^{-1}] = 2.5 \times 10^{-5}[s^{-1}]$, a rate equivalent to that of the Aβ. Those production and clearance rates are so fast that in order to achieve the desired aggregation onset time and the observed Aβ levels, the aggregation transition time must be in the order of days or else aggregation will never happen. Since we believe that aggregation transition time is in the order of years, we believe these rates represent effective clearance, modified from the dictates of estimated production rates and concentrations due to the heterogeneous character of Aβ production in the brain or by biological processes omitted from our simple model. While most aggregates and monomers are being produced and cleared at fast rates, some are in the system for a longer period of time. Those aggregates and monomers are the important ones for the aggregation process. We chose those effective rates in a way that reproduced the observed experimental data and gave slow transition times that compare well with observed onset times for disease.

The rest of the reaction parameters were adjusted in a way that the desired physical phase characteristics were reached.

**B.    Phase Diagram Exploration Method:**

We explore the dependence of several state phases of the aggregation system upon our model parameters. First we examined the state with a stable final Aβ monomer concentration that is 60% of the initial concentration, reflecting the experimental data for Aβ levels in healthy humans and those with AD [22,23,27]. Second, we determine the critical boundary between a state with no significant long time average aggregate, and those for which steady aggregation is achieved [34-36]. Initial parameters used for those phases are given in Table IV.

Once a phase state point was found, we chose two of the reaction or concentration parameters we want to draw a phase diagram for. We create the phase diagram by changing one of the parameters, and adjusting the second parameter in a way that puts us back on the phase line. We record the new point and continue to the next one. For example, in the phase state of 40% change in Aβ levels, changing one reaction parameter might cause a final Aβ level to change by only 39%, so we will need to adjust the second parameter to bring it back to 40%. Note that while along the phase line the particular chosen property is being maintained, other physical properties, like the aggregation onset time, might change.

The algorithm we developed to automate this search utilizes 4 parallel processes to follow the phase line. The phase boundary line is approximated as linear, at a small enough distance near the known state point. The linear phase line intersects a half circle curve, in the parameter space, at that small radius around that point. To find the intersection point we divide the half circle with sub-regions. Each processor runs a simulation for a sub-region. The result of the simulations indicates the region in which the solution lies. We then divide the solution region to smaller sub-regions and repeat the process until we are sufficiently close to the phase line separating distinct behavior. Once several points on the phase diagram are known, we can use a projection of the phase line to increase the step size between points. We use different random number streams for each run. The exploration process can be run in both varying parameter directions.